\documentclass[twocolumn]{aastex62}
\usepackage{amsmath}
\usepackage{CJK}
\usepackage{xcolor}

\shorttitle{Mean-Motion Resonance Pair from Microlensing}
\shortauthors{Madsen \& Zhu}

\begin{document}
\begin{CJK*}{UTF8}{gbsn}

\title{A Pair of Planets Likely in Mean-motion Resonance From Gravitational Microlensing}

\author{Sabrina Madsen}
\affil{Department of Physics, University of British Columbia Okanagan, 3333 University Way, Kelowna, BC V1V 1V7, Canada}
\affil{Canadian Institute for Theoretical Astrophysics, University of Toronto, 60 St. George Street, Toronto, ON M5S 3H8, Canada}

\author{Wei~Zhu (祝伟)}
\affil{Canadian Institute for Theoretical Astrophysics, University of Toronto, 60 St. George Street, Toronto, ON M5S 3H8, Canada}

\correspondingauthor{Wei Zhu}
\email{weizhu@cita.utoronto.ca}

\begin{abstract}
We combine statistical arguments and dynamical analysis to study the orbital configuration of the microlensing planetary system OGLE-2012-BLG-0026L. This system is known to host two massive planets, with both projected close to the Einstein ring at the time of the detection. Assuming an isotropic distribution for the orbital orientation, we find that the two planets should also be closely spaced in three-dimensional space and that the ratio of their orbital periods is almost certainly less than two. With N-body numerical integrations, we then identify two types of stable configurations: the two planets can be in first-order mean-motion resonances (MMRs) and have significant ($\gtrsim0.1$) eccentricities, or they stay out of MMRs and have nearly circular orbits. The latter is disfavored, given the absence of similar planet pairs in radial velocity (RV) observations as well as the theoretical difficulties in forming such a configuration. Therefore, the two massive planets in OGLE-2012-BLG-0026L are likely in a resonance configuration.
Our work shows that the microlensing technique, which usually only measures the projected configurations, can also probe the detailed dynamical state of multi-planet systems. We also discuss theoretical implications of measuring the multiplicity and the orbital architecture of cold planets.
\end{abstract}


\section{Introduction} \label{sec:introduction}

Gravitational microlensing can detect planetary-mass objects around stars through their perturbations on the otherwise smooth and simple \citet{Paczynski:1986} light curves \citep{Mao:1991,Gould:1992}. This technique is most sensitive to planets close to the Einstein ring, corresponding to $\sim$3 au separation for typical Galactic events (see recent reviews by \citealt{Mao:2012} and \citealt{Gaudi:2012}). Although such a feature makes it efficient in probing the cold planet population, it does place challenge in obtaining orbital information of the detected planets, as their orbital periods ($\sim$10 yr) are much longer than the duration of the microlensing event ($\sim$30 days). Consequently, for the majority of microlensing planets only the instantaneously projected separations between the planets and their hosts are constrained \citep{Gaudi:1997}.

Traditionally, there are three ways to constrain the actual three-dimensional configuration of the microlensing planetary system, but they are individually challenging and only apply to rare events. In principle, one can try to measure the radial velocities (RV) of the host star (see \citealt{Yee:2016} for a successful application in the binary star case). However, this requires the host star to be relatively bright and the planet to be fairly massive. If the planetary perturbation in the microlensing light curve persists for a significant fraction ($\gtrsim10\%$) of the orbital period, then it becomes possible to constrain the full orbit \citep{Ryu:2018}. If the planet imposes prominent features (i.e., caustic crossing) in the light curve, from which its locations at certain epochs can be precisely determined, then a much shorter perturbation can still provide meaningful constraints on the full orbit \citep{Gaudi:2008,Bennett:2010}. These latter two scenarios require special conditions, and thus they are not generally applicable.

The orbits of planets in multi-planet systems are of special interest, as they provide information on the formation and evolution of these systems. For such systems, stability analysis offers another route in constraining their orbits. A similar idea has been massively applied in the multi-planet systems found by other techniques, including radial velocity \citep[e.g.,][]{Vogt:2005,Lee:2006}, transit \citep[e.g.,][]{Lissauer:2011,Fabrycky:2014,Tamayo:2016}, and direct imaging \citep{Fabrycky:2010,Wang:2018}. In particular, the direct imaging technique commonly measures the projected positions of planets, which makes it similar to microlensing. \citet{Fabrycky:2010} applied the stability analysis to the directly imaged multi-planet system HR 8799 \citep{Marois:2008}, and found that, in order to keep the system long-term stable, the three planets should likely be in a double resonance configuration. Compared to direct imaging, which infers the planetary mass from measured flux, microlensing directly yields the planet-to-star mass ratio, which is more relevant for the dynamical analysis.

In this paper, we perform the dynamical analysis of a two-planet system detected via microlensing, OGLE-2012-BLG-0026L (hereafter OB120026L; \citealt{Han:2013}). We first summarize the known properties of this system in Section~\ref{sec:parameters}. Then, in Section~\ref{sec:orientation}, we take into account the projection effect to constrain the orbital spacing between two planets. In Section~\ref{sec:mmr} we perform N-body simulations to determine the long-term stable configurations and, based on theoretical reasonings and supporting evidence from known systems, we argue that the two planets in OB120026L are likely in one of the mean-motion resonances (MMRs). Finally, in Section~\ref{sec:discussion}, we discuss the implications from this particular system to future microlensing detections and the general demographics of exoplanets.

\section{Known Properties of OB120026L} \label{sec:parameters}

OB120026L is known to host two planets with planet-to-star mass ratios and projected separations from the host star (in units of Einstein ring radius $r_{\rm E}$)
\begin{equation}
\begin{split}
q_1 = 1.3\times10^{-4}, & ~r_{\perp,1} = 0.96 {\rm~or~} 1.03; \\
q_2 = 7.9\times10^{-4}, & ~r_{\perp,2} = 0.81 {\rm~or~} 1.25;
\end{split}
\end{equation}
respectively. Furthermore, the opening angle between the two planets is $\psi=137^\circ$ \citep{Han:2013}. The ambiguities in $r_\perp$ values come from the close-wide degeneracy \citep{Griest:1998}, as both planets were detected through their central caustic perturbations. Even so, the ratio of projected separations differs by $<30\%$ for all four solutions, corresponding to a period ratio $< 1.5$ if the actual semi-major axes are proportional to the projected separations. The fact that the two planets are not aligned (or anti-aligned) with respect to the central star also suggests that the orbital plane cannot be highly inclined, a property we will use to constrain the orbital orientation in Section~\ref{sec:orientation}.

For demonstration purposes, in the paper we only address one of the four microlensing models, namely model (B) of \citet{Han:2013}. This model has $r_{\perp,1}=1.03$ and $r_{\perp,2}=0.81$, and thus the separation ratio ($r_{\perp,\rm out}/r_{\perp,\rm in}=1.26$) is the second largest among the four models. We provide in Appendix~\ref{appendix} the results from using the other three solutions.

Although the dynamical analysis is mostly determined by the well constrained mass ratios, we notice that this system also has a well constrained stellar mass. The mass of OB120026L host star was measured to be $1.06\pm0.05~M_\odot$, by combining the lens flux and finite source effect measurements \citep{Beaulieu:2016}. This makes it a planetary system with a Sun-like host, which is not common for Galactic microlenses \citep[e.g.,][]{Zhu:2017}. The masses of the two planets are therefore $M_1=0.15~M_{\rm J}$ and $M_2=0.86~M_{\rm J}$, respectively, and the Einstein ring radius $r_{\rm E}=4~$au. The well determined physical properties of this system make it possible to connect with known exoplanet demographics around Sun-like stars, as we will discuss in Section~\ref{sec:discussion}.

\section{Orbital Orientation of OB120026L} \label{sec:orientation}

The microlensing light curve provides stringent constraints ($\lesssim 3\%$) on the mass ratio $q$ and the projected positions (i.e., dimensionless separation $s$ and azimuthal separation $\psi$),
\footnote{The uncertainties on masses and absolute projected separations are relatively large ($\sim10\%$). These are dominated by the uncertainties on the lens mass measurement \citep{Beaulieu:2016}. However, the dynamics of the system can be well described by the dimensionless parameters $q$, $s$ and $\psi$, and the absolute scale only has marginal effect.}
but it has nearly zero constraint on the projected velocities as well as positions and velocities along the line-of-sight direction \citep{Han:2013}. In this section, we explore the impact of these unknown parameters on the relative positions between two planets in the orbital plane.

The projected position of the planet $j$ ($j=1$ or 2) relative to the host is given by
\begin{equation}
\begin{pmatrix}
x_j \\ y_j
\end{pmatrix} = r_j
\begin{pmatrix}
\cos{\Omega} \cos{u_j} - \sin{\Omega} \sin{u_j} \cos{i} \\
\sin{\Omega} \cos{u_j} + \cos{\Omega} \sin{u_j} \cos{i}
\end{pmatrix} \ ,
\end{equation}
where $\Omega$ is the longitude of ascending node, $i$ is the orbital inclination, $u$ is the argument of latitude (i.e., the sum of the argument of periapsis $\omega$ and the true anomaly $f$), and $r$ is the radial separation between the planet and the star. We define the $z$-axis to be the direction toward the observer, and the $x$-axis to be the direction toward the projected position of the lower-mass planet (see Figure~2 of \citealt{Han:2013}). We assume that the two planets are coplanar. The deviation from this assumption will be addressed at the end of this section. 

\begin{figure*}
\epsscale{1.1}
\plotone{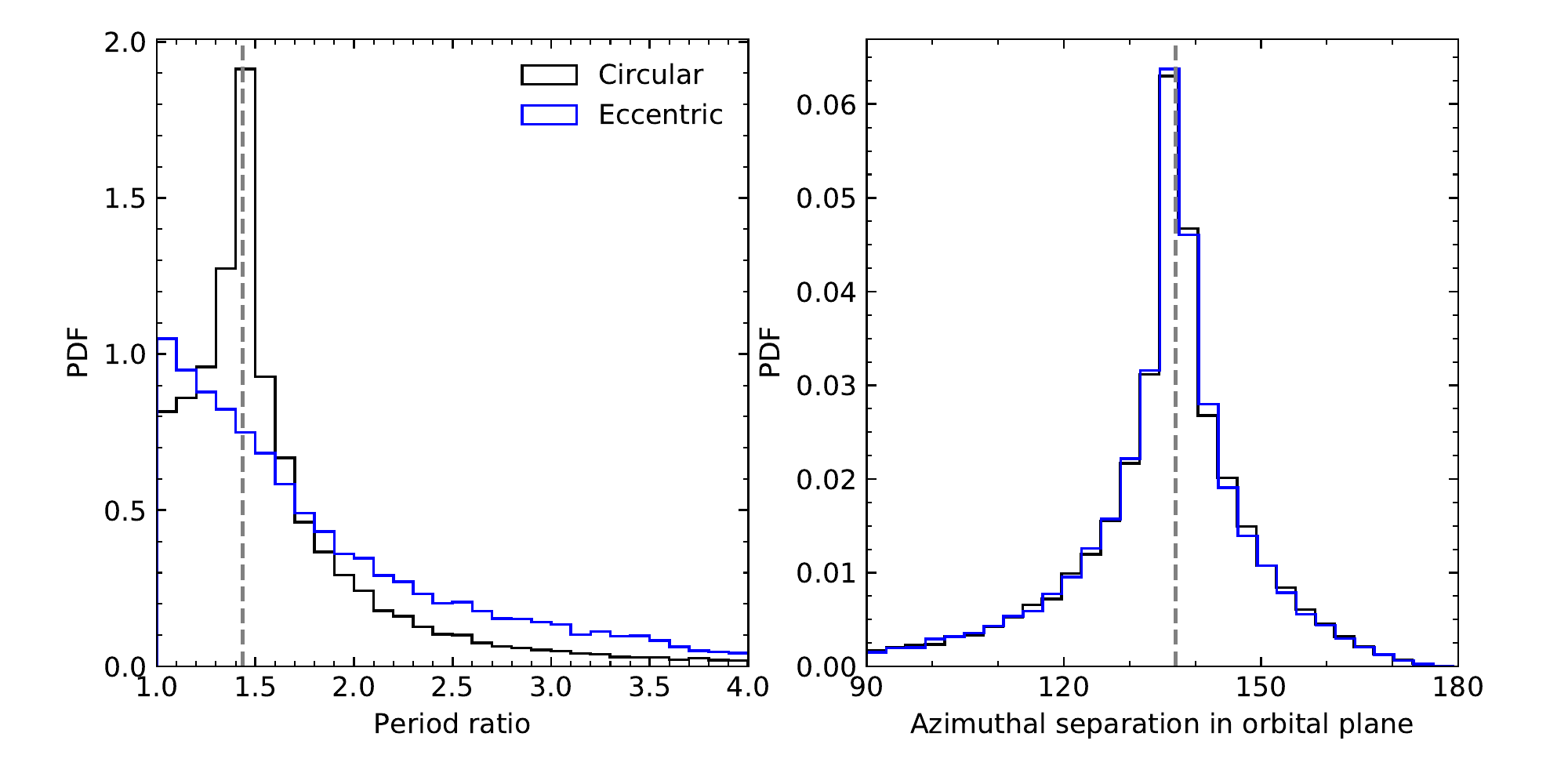}
\caption{Probability distribution functions of the the period ratio (left panel) and the azimuthal separation between the two planets in the orbital plane (right panel). In the eccentric case, we fix the eccentricities of both planets to $0.3$. The vertical dashed lines indicate the observed values, which is $(r_{\perp,\rm out}/r_{\perp,\rm in})^{3/2}$ in the left panel and the value of $\psi$ in the right panel.
\label{fig:posteriors}}
\end{figure*}

We are mostly interested in the probability distribution of the period ratio between the two planets. To reach this goal, we first randomly choose $i$ from an isotropic distribution ($\propto \sin{i}$) and $\Omega$ from a uniform distribution between $0$ and $2\pi$. We then solve the above set of equations for $r_j$ and $u_j$ ($j=1,~2$). For circular orbits, the semi-major axis, $a$, is given by $r$ and then the period ratio can be derived from the semi-major axis ratio. We always report the outer-to-inner period ratio. To account for the commonly adopted log-flat distribution in $a$, we weight the derived period ratio by $1/(a_1 a_2)$. From $50,000$ random sets of $i$ and $\Omega$ we construct the probability distribution of the period ratio as well as the distribution of the azimuthal separation between the two planets in the orbital plane (i.e., $u_2-u_1$), both of which are illustrated in Figure~\ref{fig:posteriors}. The two planets likely have small period ratios even after the correction of the projection effect. Although the period ratio distribution has a long tail toward larger values, it is unlikely ($<14\%$) for the period ratio to exceed 2. There is also a non-negligible probability that the two planets are in a super compact (period ratio close to unity) configuration, the majority of which is dynamically unstable. This is further explored in Section~\ref{sec:mmr}. The observed azimuthal separation $\psi$ also serves as a good proxy for the azimuthal separation between the two planets in the orbital plane.

Introducing eccentric orbits only changes both probability distributions at a quantitative level. To demonstrate this, we carry out the following exercise. For each orientation of the orbital plane, we also introduce to each planet an eccentricity vector with a fixed amplitude $e=0.3$ but a random direction $\omega$. The semi-major axis is then given by $a=r[1+e\cos(u-\omega)]/(1-e^2)$, and the rest procedure is the same as in the circular case. The resulting probability distributions are also shown in Figure~\ref{fig:posteriors}. Even with such extreme eccentricities, the key feature remains: the period ratio is most likely smaller than 2.

We now discuss the impact of deviations from coplanar orbits. If the orbit of the apparently inner planet (planet 2) is more inclined with respected to the other one, the period ratio becomes even smaller than that derived under the coplanarity assumption, unless the mutual inclination becomes larger than $\cos^{-1}{[r_{\perp,2}/r_{\perp,1}]}=38^\circ$. After that the apparently inner planet becomes the outer one. Such a configuration becomes a hierarchical system, which we defer to Section~\ref{sec:discussion} for further discussion. Inclining the orbit of the apparently outer planet (planet 1) would increase the period ratio, but unless the outer planet is significantly inclined, the period ratio derived from the projected separations serves as a reasonably good approximation of the actual period ratio.
\footnote{Retrograde orbits can in principle produce a more stable system than the prograde orbits \citep{Nesvorny:2003}. However, such a configuration is considered unlikely and therefore not discussed in this paper.}

To summarize, the closely spaced configuration of the two-planet system OB120026L in the projected plane strongly suggests that the two planets are also closely spaced in 3D space.

\section{Two Planets Likely in Mean-Motion Resonances} \label{sec:mmr}

\begin{figure*}
\epsscale{1.1}
\plotone{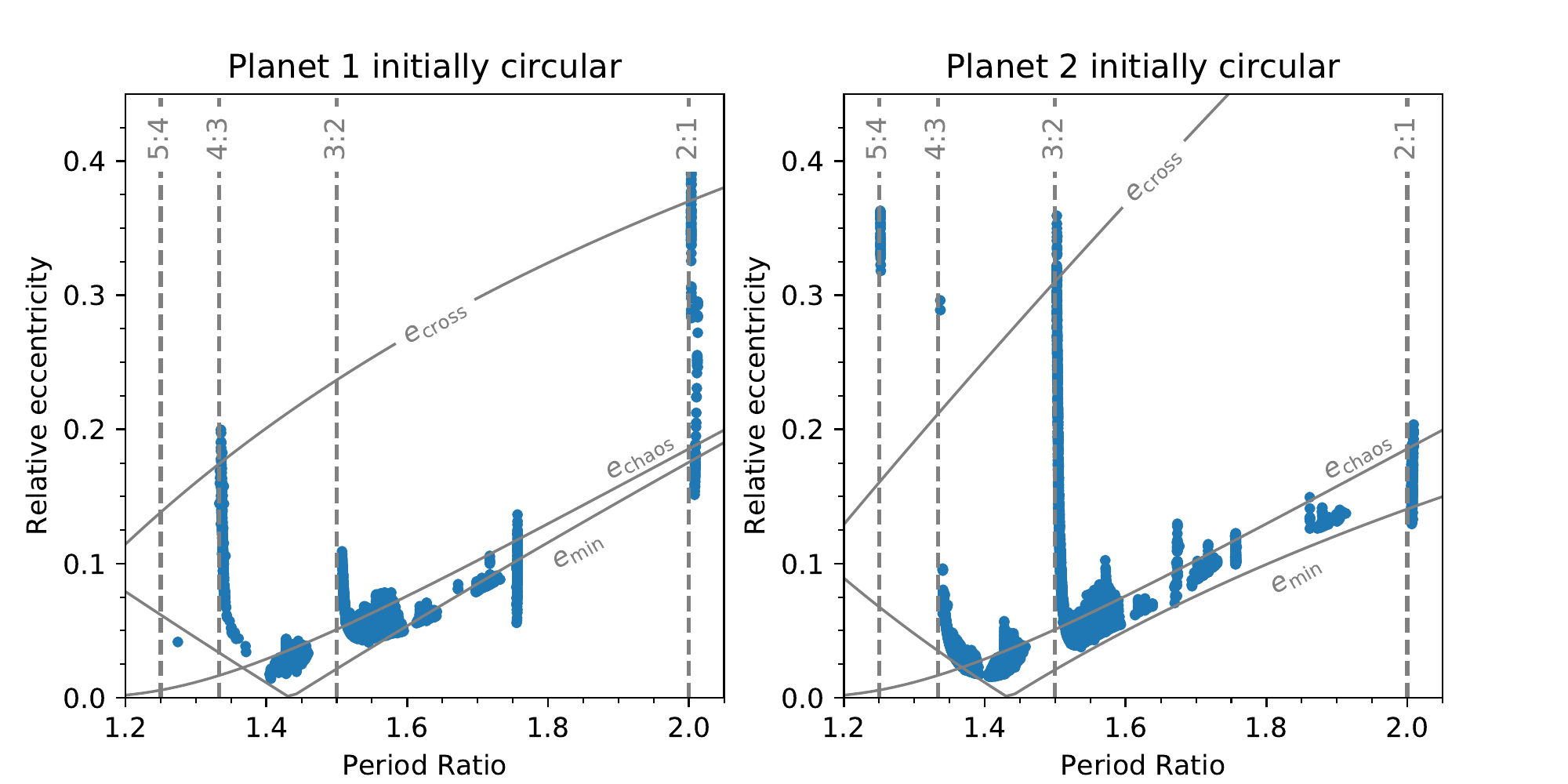}
\caption{Relative eccentricities (Equation~(\ref{eqn:relative_ecc})) and period ratios of all regular configurations, for the case in which planet 1 is initially circular (left panel) and the case in which planet 2 is initially circular (right panel). What is shown here are the median values of librating eccentricities and period ratios, computed from 2000 samplings over the 3000-orbit integration. Vertical dashed lines are the integer ratios that correspond to first-order MMRs. In each panel we also show with solid curves three types of threshold eccentricities: $e_{\rm cross}$ is the threshold for orbit crossing, $e_{\rm chaos}$ is the threshold for onset of chaos (given by Equation (19) of \citealt{Hadden:2018}), and $e_{\rm min}$ the minimum relative eccentricity for the observed projected separations to reach a certain period ratio. Due to librations on secular timescales and our limited integration time, some data points appear below the $e_{\rm min}$ curve. The observed separation ratio corresponds to a period ratio of $1.43$, which is the bottom of the V-shaped $e_{\rm min}$ curve.
\label{fig:mmrs}}
\end{figure*}

Here we investigate the impact of dynamical instability on the system configuration. Because the two planets are closely spaced, mean motion resonances (MMRs), which are not captured by the stability criteria either derived analytically \citep[e.g.,][]{Deck:2013,Hadden:2018} or empirically \citep[e.g.,][]{Petrovich:2015}, make a non-negligible contribution to all stable configurations. Therefore, we use N-body numerical integrations to determine the stability of any given configuration. These are done with the \texttt{REBOUND} \citep{Rein:2012} code with the WHFast integrator \citep{Rein:2015}, which is a fast and accurate implementation of the \citet{Wisdom:1991} symplectic algorithm.

We further assume that the planetary system is face-on ($i=0$). 
Because the spacing between two planets, which is most relevant for this dynamical analysis, is very small even when different orientations of the orbital plane are taken into account (see Section~\ref{sec:orientation}), our result below is not affected by the choice of this initial condition.

In principle, the two detected planets can both have eccentric orbits, but we consider eccentric orbits for two planets separately and assume initially circular orbit for the other planet. In each case, we run $10^5$ simulations. The initial state of the planet in a circular orbit is fully determined by the known $(x,~y)$ positions. For the eccentric orbit, we first randomly choose eccentricity, $e$, uniformly between 0 and 0.5 and mean anomaly, $l$, uniformly between 0 and $2\pi$, and then solve Kepler's equation for eccentric anomaly, $E$. Together with the known $(x,~y)$ coordinates, these parameters yield the semi-major axis, $a$, and the argument of periapsis, $\omega$, and thus the initial state of this eccentric planet is also fully determined. For each simulation, we run up to 3000 orbits of the longer-period planet, with a step size of one 100th of the smaller orbital period. To distinguish between regular and chaotic orbits, we use the Mean Exponential Growth of Nearby Orbits (MEGNO) chaos indicator \citep{Cincotta:2003} built into \texttt{REBOUND} and use MEGNO=2 as the threshold.
\footnote{This is a conservative threshold since the MEGNO value of a regular orbit can be slightly larger than 2 for a finite integration time. However, our result is not sensitive to the choice of the MEGNO threshold, as similar simulations have shown (e.g., Figure 6 of \citealt{Hadden:2018}).}
The system is also determined to be unstable if either of the planets is too far ($>20$ au) away from or too close ($<0.01$ au) to the central star. This last criterion is introduced in order to make the computations more efficient. For a subset (139) of the stable simulations, we integrate up to $10^5$ orbits. A large fraction of these integrations lead to unstable orbits, but there is no systematic preference over any specific initial condition (low-$e$, non-resonant or high-$e$, resonant). Therefore, the results we report below are not affected by the integration runtime.

We show the relative eccentricities and period ratios of the regular configurations (MEGNO$\le2$) in Figure~\ref{fig:mmrs}. Here the relative eccentricity is defined as
\begin{equation} \label{eqn:relative_ecc}
Z \equiv \frac{| e_2 e^{i\varpi_2} - e_1 e^{i\varpi_1} |}{\sqrt{2}}\ ,
\end{equation}
where $\varpi_1$ and $\varpi_2$ are the longitudes of periapsis of the two planets, respectively \citep{Hadden:2018}. Given our choices of the initial condition, this is approximately the initial eccentricity of the eccentric planet. As Figure~\ref{fig:mmrs} shows, there are two different types of regular configurations regardless of which planet is set eccentric initially: planets have nearly circular ($e\lesssim0.1$) orbits, or planets have significant eccentricities and can be in one of the first-order MMRs. In fact, as seen in Figure~\ref{fig:mmrs}, there are regular configurations under which the two planets have crossing orbits, and it is MMR that protects the system from instability.

\begin{figure}
\epsscale{1.1}
\plotone{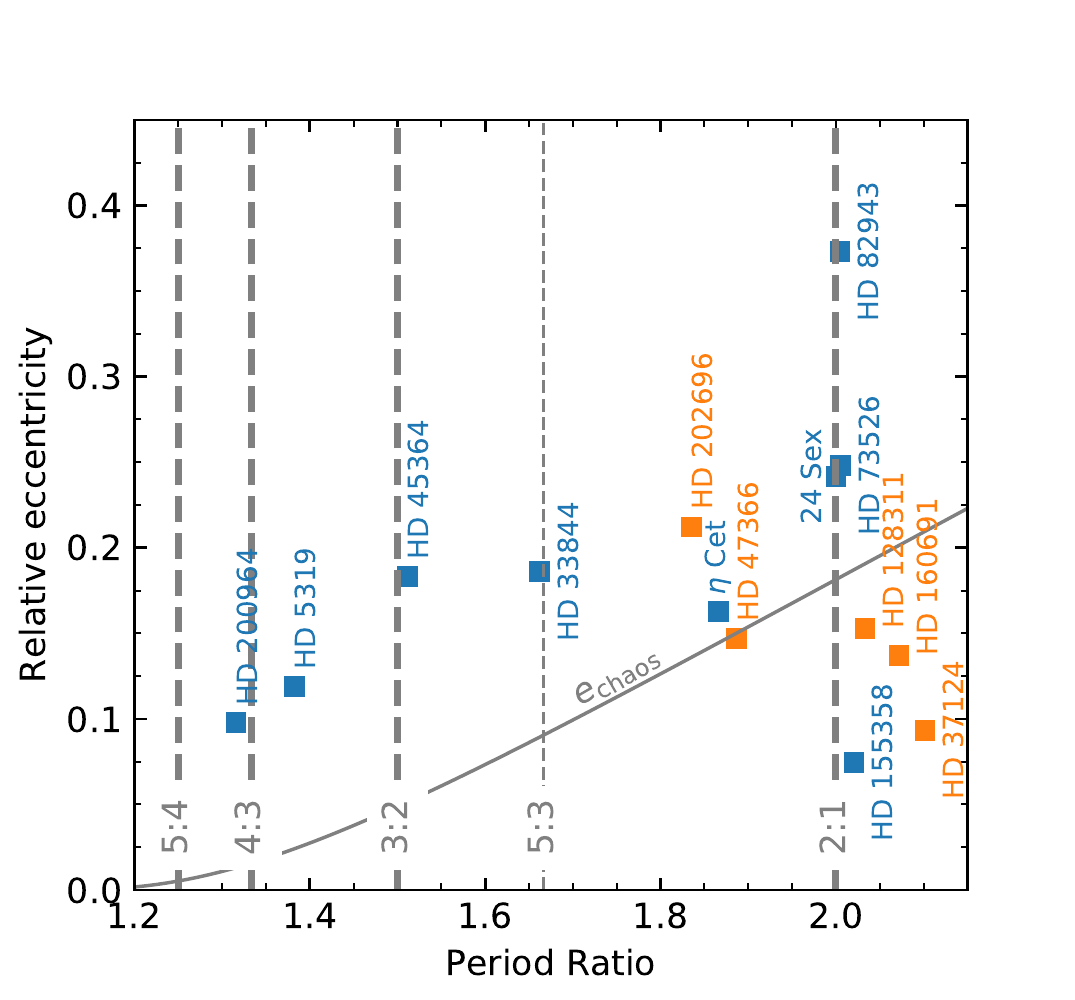}
\caption{Known massive ($>0.15~M_{\rm J}$) planet pairs with $P_{\rm out}/P_{\rm in}<2.2$. The y-axis shows the relative eccentricity computed from the reported eccentricities and argument of periapses. The threshold for onset of chaos, $e_{\rm chaos}$, is computed for $q_{\rm tot}=10^{-3}$. Systems colored in blue are those for which resonant configurations are preferred, whereas those colored in orange are systems for which resonance is disfavored but remains possible. See Table~\ref{tab:rv-pairs} for the details of individual systems.
\label{fig:known_pairs}}
\end{figure}

Although both types of configurations are allowed from the stability point of view, the MMR configurations are preferred once the formation history of such a planet pair is taken into account. Figure~\ref{fig:known_pairs} shows the selected planet pairs from NASA Exoplanet Archive \citep{Akeson:2013}.
\footnote{We also included $\eta$ Cet and HD 202696. Both systems have planet pairs that meet our criteria.}
Parameters of individual systems are provided in Table~\ref{tab:rv-pairs}. Similar to the two planets in OB120026L, these planets are relatively massive ($>0.15~M_{\rm J}$) and well-separated ($P>100~$d) from their hosts. In the period ratio range $1<P_{\rm out}/P_{\rm in}<2.2$, the majority of the planet pairs have period ratios close to commensurabilities, with half close to the 2:1 ratio. Dynamical analyses have suggested the existence of MMRs in most cases, in particular if the period ratio is $<2$.

The lack of giant planet pairs in closely spaced but non-MMR configurations in Figure~\ref{fig:known_pairs} is also well understood theoretically. When protoplanets grow, orbital repulsion keeps the separation between neighboring planets wider than 5--10 times the mutual Hill radius of the protoplanets \citep{Kokubo:1998}. Here the mutual Hill radius is given by
\begin{equation}
r_{\rm H} \equiv \frac{a_{\rm in}+a_{\rm out}}{2} \left(\frac{q_{\rm in}+q_{\rm out}}{3}\right)^{1/3}\ .
\end{equation}
For OB120026L, this means that the two planets should start with at least $P_{\rm out}/P_{\rm in}>1.6$ and more likely $P_{\rm out}/P_{\rm in}>2.2$. The initial orbits are also nearly circular ($e < 0.1$), due to the interactions between protoplanets and the embedded disk. To produce planet pairs in smaller period ratios, convergent migration is required, and together with eccentricity damping this usually leads to planet pairs trapped in MMRs \citep{Lee:2002}. It is difficult to form closely spaced massive planet pairs that are not in MMRs without fine-tuning the conditions.

The above theoretical arguments and the lack of planet pairs outside MMRs in small period ratios in Figure~\ref{fig:known_pairs} both suggest that the MMR configurations are preferred for the two-planet system OB120026L.

\section{Discussion} \label{sec:discussion}

We combine statistical arguments and dynamical analysis to study the orbital configuration of the two-planet system OB120026L that was found through gravitational microlensing \citep{Han:2013}. The two massive planets, one with $0.86~M_{\rm J}$ and the other with $0.15~M_{\rm J}$ \citep{Beaulieu:2016}, were found to be closely spaced in the sky-projected plane, with the ratio of the projected separations $<1.3$. Under the reasonable assumptions that the orbital orientation follows an isotropic distribution and that the mutual inclinations are typically small, this observed configuration strongly suggests that the two planets should also be closely spaced in the 3D space, and that the period ratio is unlikely to exceed two (Figure~\ref{fig:posteriors}).

We then study the eccentric orbits with numerical integrations and find two types of stable configurations. The two planets can be in MMRs and have significant ($e\gtrsim0.1)$ eccentricities, or they have nearly circular ($e\lesssim0.1$) orbits and stay out of nominal MMRs. The out-of-MMR configurations are disfavored, given the absence of similar planet pairs from RV observations and the theoretical difficulties in forming such configurations. Therefore, the two massive planets in OB120026L are probably in MMR.

The above conclusion builds on the assumption that the two planets in OB120026L have nearly coplanar ($\lesssim 38^\circ$, see Section~\ref{sec:orientation}) orbits. Given our understanding of the architecture of planetary systems from both observations and theories, this is a fair assumption, although other scenarios cannot be entirely ruled out. For example, a hierarchical, highly mutually inclined two-planet system can also produce the observed configuration by chance. We consider such a scenario unlikely, given the small \textit{a priori} probability for the two planets in such a hierarchical system to have similar projected separations.
\footnote{Assuming isotropic orientation, the chance for a planet with semi-major axis $a$ to have a projected separation $r_\perp$ is simply $r_\perp/a$. With $r_\perp\approx4$ au, this is $10\% (a/{\rm 40~au})^{-1}$. The two planets in such hierarchical systems will likely have dynamical interactions (i.e., Lidov-Kozai libration \citealt{Lidov:1962,Kozai:1962}), the inclusion of which will further decrease the probability.}
In principle, this low by-chance projection probability can be compensated by a much higher occurrence rate. This, if true, will lead to an enhanced rate of microlensing detections in which a normal planetary event is followed (or preceded) by a short single-lens event due to the distant planet.
\footnote{See OGLE-2008-BLG-092 \citep{Poleski:2014} for an example, although in this case the distant companion is a star.}
Future microlensing observations will be able to test whether this is true or not.

Our work shows that the microlensing technique, which usually only measures the projected configuration, can also probe the detailed dynamical state of the multi-planet system. The two features of OB120026L that are key to our conclusion are that 1) the two planets are well separated azimuthally, and 2) they are both close to the Einstein ring. For a comparison, the first two-planet system, OGLE-2006-BLG-109L, has two planets almost aligned ($\sim13^\circ$) with the central star and the projected separation ratio $\sim1.7$ \citep{Gaudi:2008,Bennett:2010}, making our method inapplicable. So is OB120026L-type configuration rare in microlensing detections? The detections of multiple planets in the same system are largely independent \citep{Gaudi:1998,Zhu:2014a,Zhu:2014b,Shin:2015}. Hence, it is more likely to detect two-planet systems with both planets close to the Einstein ring, given that microlensing sensitivity is a steep function of the distance from the Einstein ring. Furthermore, there is no obvious preference to any particular azimuthal separation in two-planet detections, as shown in previous simulations \citep{Zhu:2014a}. Therefore, we expect that a significant fraction of future two-planet systems (and likely higher multiples) should have configurations similar to OB120026L, and thus the method we developed here will be generally applicable.

The orbital architecture and occurrence rate of planetary systems like OB120026L, which hosts both a cold Jupiter ($q\sim10^{-3}$) and a cold Neptune ($q\sim10^{-4}$), are important for our better understanding of the planet formation process. Capturing planets into MMRs requires some dissipative migration through interactions with the gas disk or planetesimals \citep[e.g.,][]{Lee:2002}, so the fraction of planet pairs in MMRs is an important proxy for understanding how planetary architectures are set during formation. Ground-based RV observations find that about 1/3 of well-characterized multi-planet systems contain planet pairs with apparent low-order period commensurability \citep{Wright:2011}. Such a high concentration cannot be explained by random fluctuations, but is consistent with the expectation from migration theory \citep{Goldreich:1980}. By contrast, the \emph{Kepler} mission found that the close-in small planets do not show a strong preference for MMRs \citep{Lissauer:2011}, suggesting that migration, or at least large scale migration, might not have happened for these planets \citep{Petrovich:2013}. These two types of planets differ in both mass and orbital separation, and therefore the cold Neptunes that microlensing is sensitive to \citep{Gould:2006}, which have similar separations to the RV cold giants but similar masses to the \emph{Kepler} super Earths, convey important clues on what drives the different architectures. 

It is now known that cold Jupiters are almost certainly accompanied by inner small planets \citep{ZhuWu:2018}, but how often do they also have cold Neptune companions? While this question can serve a test of planet formation theories, it is also relevant for understanding the prevalence of the solar system-like architecture. By the end of 2018, there were 35 microlensing planetary systems with $q>4\times10^{-4}$ listed on NASA Exoplanet Archive \citep{Akeson:2013}, out of which two were two-planet systems with cold Neptunes.
\footnote{There is also a candidate two-planet system OGLE-2014-BLG-1722, but both planets have $q>4\times10^{-4}$ \citep{Suzuki:2018}.}
If one simply takes the detection efficiency of an additional planet from the simulation ($\sim5\%$, \citealt{Zhu:2014a}), these numbers seem to suggest that all cold Jupiters can be accompanied by cold Neptunes. However, the above result has to be taken with caution because of the mismatch between observations and the simulation. Both two-planet systems were detected in high-magnification events that received intensive follow-up observations \citep[e.g.,][]{Gould:2010}, whereas the simulation done by \citet{Zhu:2014a} assumed an observing strategy with a consistent cadence (10 min) similar to the Korean Microlensing Telescope Network \citep[KMTNet,][]{Kim:2016}. Therefore, more detailed works and larger samples are needed in order to settle this issue.

It is also interesting to notice that the first two multi-planet systems from microlensing both contain one planet with intermediate mass ($30-100~M_\oplus$): OGLE-2006-BLG-109c has $86~M_\oplus$ \citep{Bennett:2010}, and OGLE-2012-BLG-0026c has $46~M_\oplus$ \citep{Beaulieu:2016}.
\footnote{Both events have finite-source effect, microlensing parallax, and lens flux measurement, and thus the mass measurements are fairly secure.}
In the standard core accretion scenario \citep{Pollack:1996}, hydrostatic gas accretion can no longer be maintained once the gaseous envelop doubles the planetary core mass ($\sim10~M_\oplus$), and subsequent run-away gas accretion pushes the total mass rapidly to the gas giant regime ($\gtrsim100~M_\oplus$). Therefore, the population synthesis models based on the core accretion theory predicted that planets with masses in the intermediate mass range should be very rare \citep{IdaLin:2004,Mordasini:2009}. This is not supported by microlensing observations, which have found that the transition from Neptune mass to Jupiter mass is rather smooth \citep{Suzuki:2016}. One possibility to reconcile the theory and observations, as suggested by the first two multi-planet systems, is that the multiplicity rate may be higher than model predicts, and that competitions for gas material starve some of the embryos from fully growing into giants. Such a scenario is particularly possible for OB120026L, since in this system the two planets are so closely spaced. Future microlensing studies on the multiplicity rate can confirm whether this explanation stands or not.

\acknowledgements
We would like to thank Cristobal Petrovich, Yanqin Wu, Dan Tamayo, and Mohamad Ali-dib for useful discussions, and Andy Gould for comments on an earlier version of the paper. We also thank the anonymous referee for comments and suggestions that helped improve the manuscript.
W.Z. also thanks Dan Fabrycky for some early discussions. 
This work is the result of the 2018 Summer Undergraduate Research Program (SURP) in astronomy \& astrophysics at the University of Toronto.
W.Z. was supported by the Beatrice and Vincent Tremaine Fellowship at CITA.


\appendix
\section{Results from other Solutions} \label{appendix}

\begin{figure*}
\epsscale{1.1}
\plotone{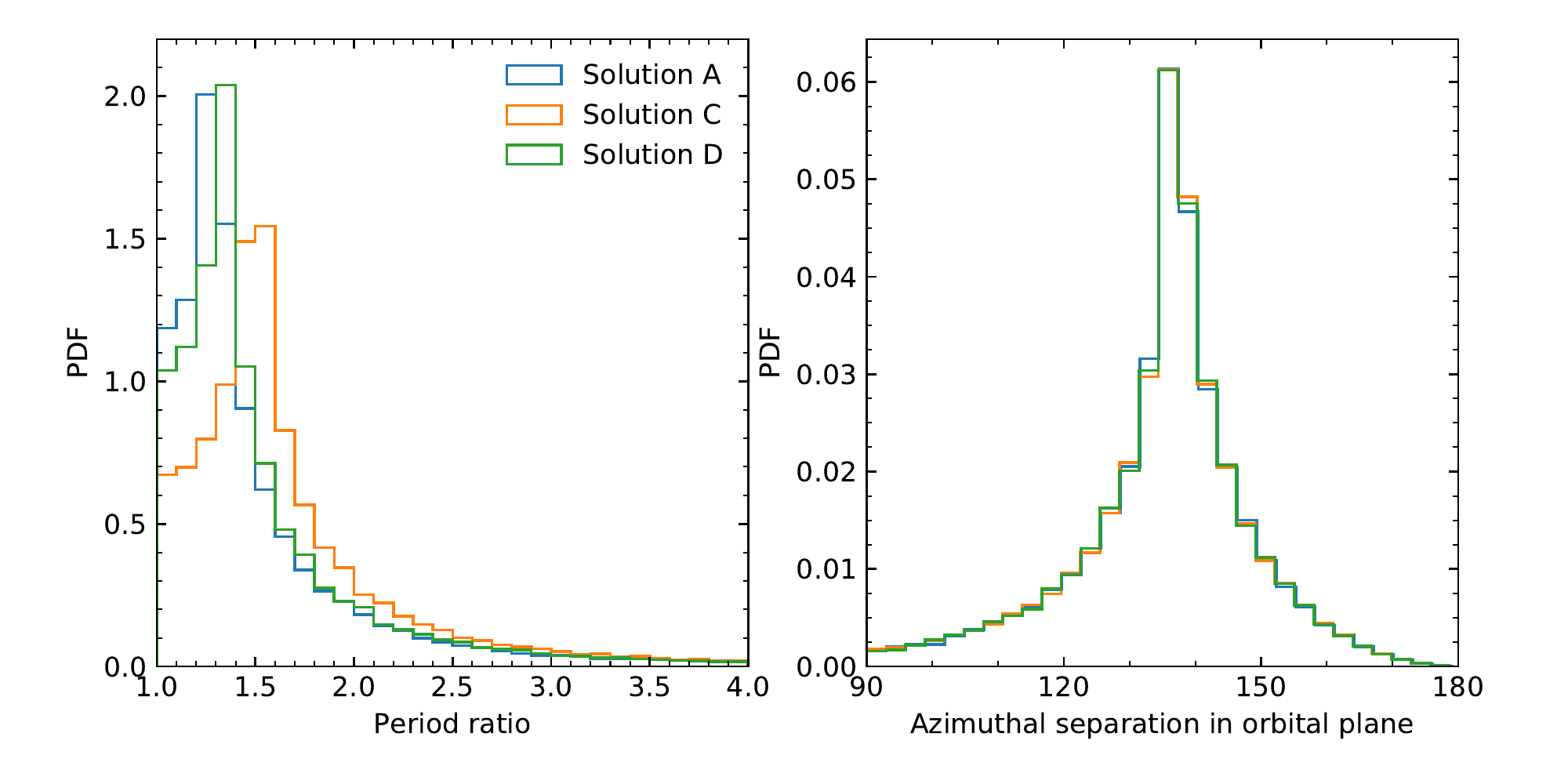}
\caption{Probability distribution functions of the the period ratio (left panel) and the azimuthal separation between the two planets in the orbital plane (right panel) for the other three solutions in \citet{Han:2013}.
\label{fig:other-posteriors}}
\end{figure*}

\begin{figure*}
\epsscale{1.1}
\plotone{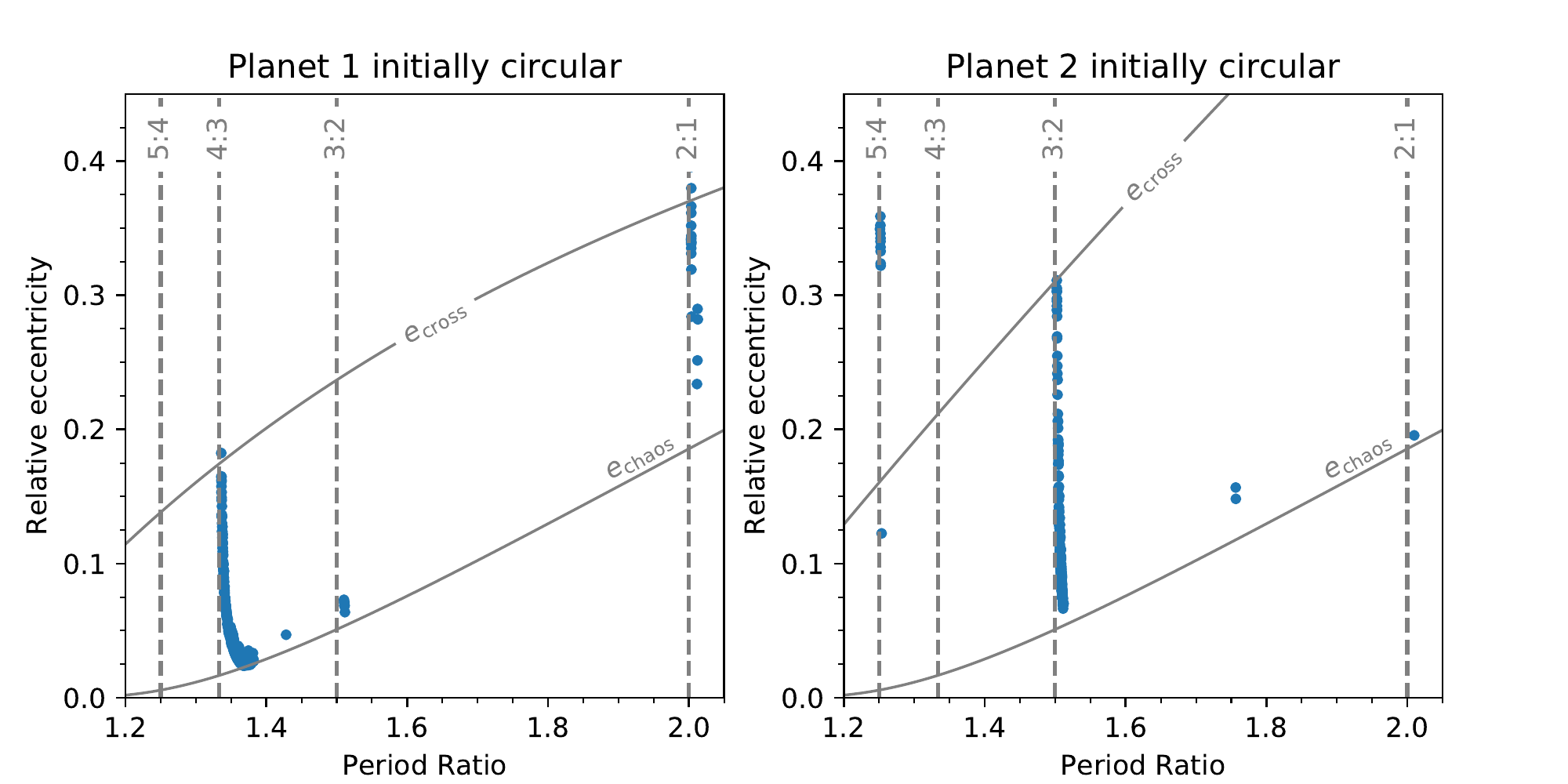}
\caption{Similar to Figure~\ref{fig:mmrs} but for model (A) of \citet{Han:2013}. In this case the inner planet (i.e., planet 2) is more massive, $r_{\perp,1}=0.96$, and $r_{\perp,2}=0.81$. The face-on and circular configuration is unstable because the two planets are so closely spaced.
\label{fig:model-A}}
\end{figure*}

\begin{figure*}
\epsscale{1.1}
\plotone{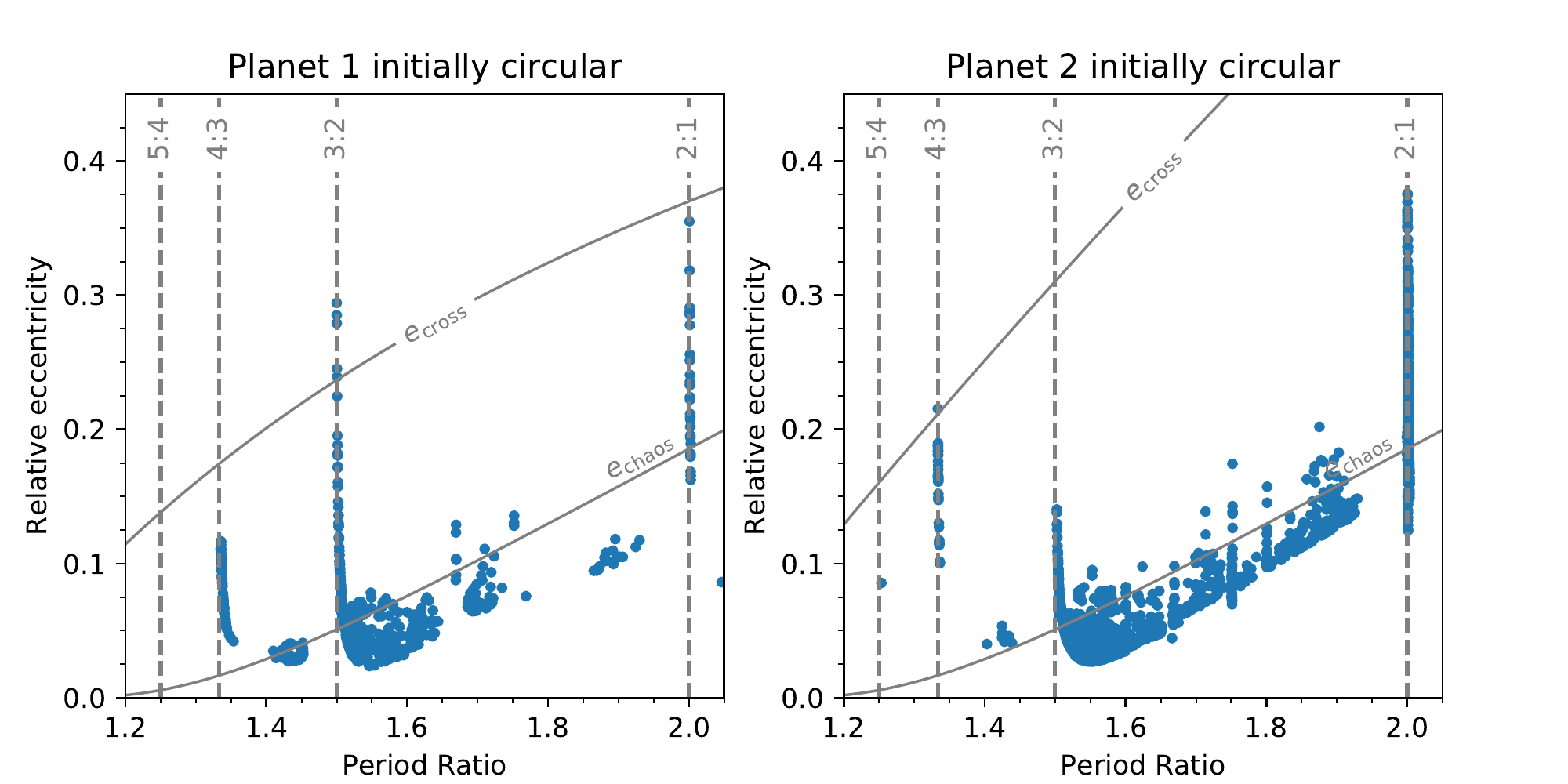}
\caption{Similar to Figure~\ref{fig:mmrs} but for model (C) of \citet{Han:2013}. In this case the outer planet (i.e., planet 2) is more massive, $r_{\perp,1}=0.96$, and $r_{\perp,2}=1.26$.
\label{fig:model-C}}
\end{figure*}

\begin{figure*}
\epsscale{1.1}
\plotone{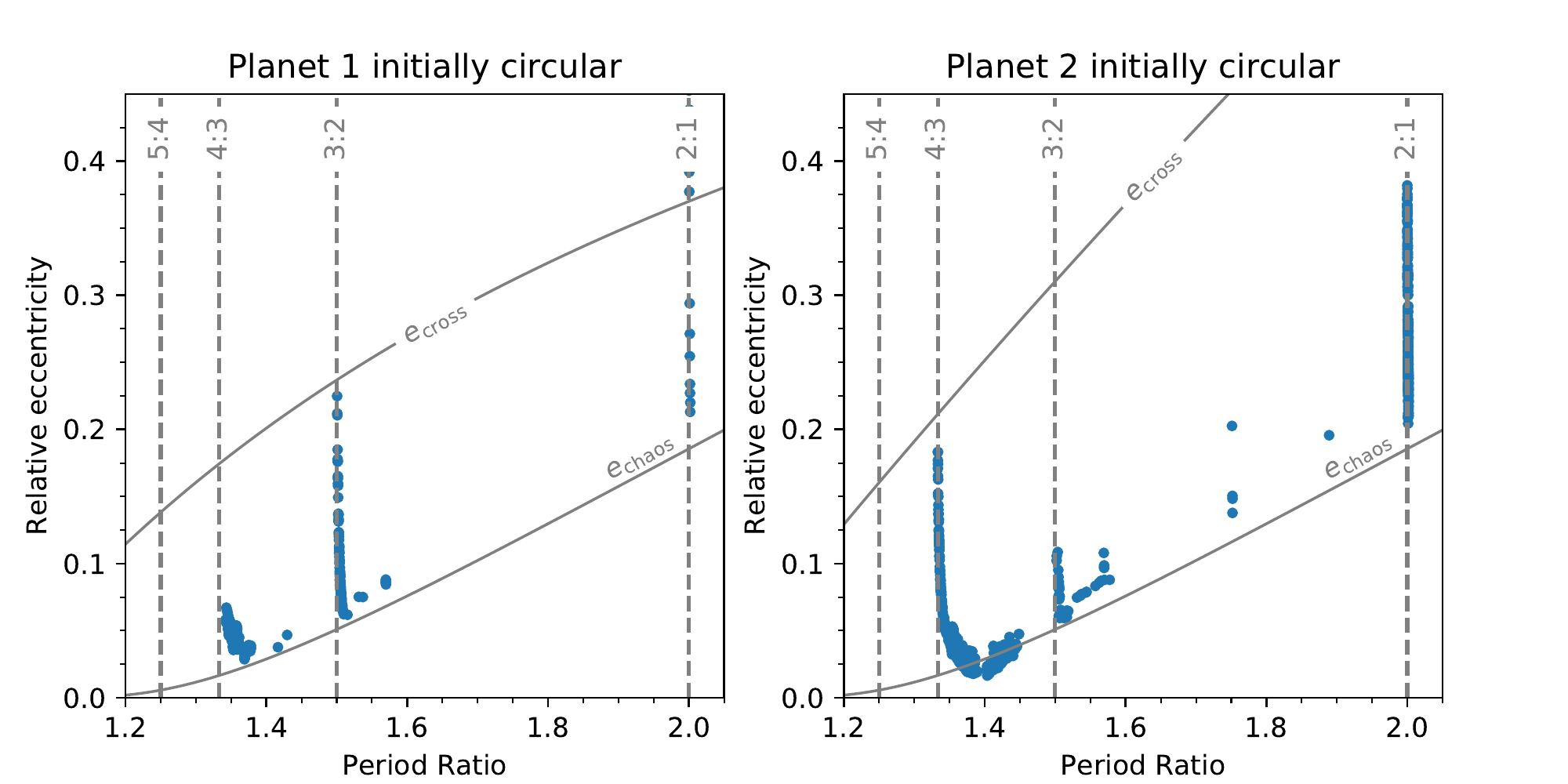}
\caption{Similar to Figure~\ref{fig:mmrs} but for model (D) of \citet{Han:2013}. In this case the outer planet (i.e., planet 2) is more massive, $r_{\perp,1}=1.03$, and $r_{\perp,2}=1.25$.
\label{fig:model-D}}
\end{figure*}

Our main results do not change even when other allowed microlensing solutions are used. Here we show the results from the other three allowed solutions given in \citet{Han:2013}. Figure~\ref{fig:other-posteriors} shows the probability distributions of the period ratio and the azimuthal separation in the orbital plane (i.e., before projection effect). Figures~\ref{fig:model-A}, \ref{fig:model-C}, and \ref{fig:model-D} show the stable configurations out of the N-body integrations. For each case we have $10^4$ realizations.

\section{Packed Massive Planet Pairs from RV} \label{sec:rv-pairs}

We list in Table~\ref{tab:rv-pairs} the information of RV planet pairs used in Figure~\ref{fig:known_pairs}. Some comments on the dynamical analysis are also provided.

\begin{deluxetable}{lcllclc}
\tablecaption{Massive planet pairs with period ratio $<2.2$.
\label{tab:rv-pairs}}
\tablehead{
\colhead{Host name} & \colhead{$M_\star/M_\odot$} & \colhead{$(m\sin{i},~P,~e,~\omega)_{\rm in}$\tablenotemark{a}} & \colhead{$(m\sin{i},~P,~e,~\omega)_{\rm out}$\tablenotemark{a}} & \colhead{MMR\tablenotemark{b}?} & \colhead{Reference} & \colhead{Comments}}
\startdata
HD 200964 & $1.4$ & $(1.9,~631,~0.11,~223)$ & $(1.3,~829,~0.11,~301)$ & Yes & \citet{Johnson:2011} & (1) \\
HD 5319 & $1.5$ & $(1.8,~641,~0.02,~97)$ & $(1.2,~886,~0.15,~252)$ & Yes & \citet{Giguere:2015} & (2) \\
HD 45364 & $0.8$ & $(0.19,~227,~0.17,~163)$ & $(0.66,~343,~0.10,~7.4)$ & Yes & \citet{Correia:2009} & (3) \\
HD 33844 & $1.7$ & $(2.0,~551,~0.15,~211)$ & $(1.8,~916,~0.13,~71)$ & Yes & \citet{Wittenmyer:2016} & (4) \\
HD 47366 & $1.8$ & $(1.8,~363,~0.09,~100)$ & $(1.9,~685,~0.28,~132)$ & Possible & \citet{Sato:2016} & (5) \\
24 Sex & $1.5$ & $(1.6,~455,~0.18,~227)$ & $(1.4,~910,~0.41,~172)$ & Yes & \citet{Johnson:2011} & (6) \\
$\eta$ Cet & $1.7$ & $(2.6,~404,~0.13,~251)$ & $(3.3,~752,~0.10,~68)$ & Yes & \citet{Trifonov:2014} & (7) \\
HD 202696 & $1.9$ & $(1.9,~521,~0.06,~259)$ & $(2.0,~956,~0.26,~129)$ & Possible & \citet{Trifonov:2019} & (8) \\
HD 73526 & $1.1$ & $(2.3,~189,~0.29,~196)$ & $(2.3,~379,~0.28,~272)$ & Yes & \citet{Wittenmyer:2014} & (9) \\
HD 82943 & $1.2$ & $(1.9,~220,~0.37,~117)$ & $(1.7,~441,~0.16,~300)$ & Yes & \citet{Baluev:2014} & (10) \\
HD 155358 & $1.2$ & $(1.0,~194,~0.17,~143)$ & $(0.82,~392,~0.16,~180)$ & Yes & \citet{Robertson:2012} & (11) \\
HD 128311 & $0.8$ & $(1.8,~453,~0.30,~58)$ & $(3.1,~922,~0.16,~15)$ & Possible & \citet{McArthur:2014} & (12) \\
HD 160691 & $1.1$ & $(0.52,~310,~0.07,~190)$ & $(1.1,~643,~0.13,~22)$ & Possible & \citet{Pepe:2007} & (13) \\
HD 37124 & $0.9$ & $(0.65,~885,~0.13,~53)$ & $(0.70,~1862,~0.16,~0)$ & Possible & \citet{Wright:2011} & (14) \\
\enddata
\tablecomments{
(1) Two planets are almost certainly in 4:3 MMR (see also \citealt{Wittenmyer:2012}).
(2) Two planets are most likely in 4:3 MMR.
(3) Two planets are most likely in 3:2 MMR.
(4) Two planets are most likely in 5:3 MMR.
(5) Low-$e$, non-resonant solutions are preferred, but resonant configurations are also possible (see also \citealt{Marshall:2019}).
(6) Two planets are almost certainly in 2:1 MMR (see also \citealt{Wittenmyer:2012}).
(7) Compared to non-resonant solutions, 2:1 resonant solutions are statistically preferred \citep{Trifonov:2014}.
(8) The two planets are likely not in MMR, but 2:1 MMR is not ruled out \citep{Trifonov:2019}.
(9) The two planets are almost certainly in 2:1 MMR (see also \citealt{Tinney:2006}).
(10) The two planets are almost certainly in 2:1 MMR (see also \citealt{Lee:2006}).
(11) The two planets are most likely in 2:1 MMR, although stable non-resonant solutions also exist \citep{Robertson:2012}.
(12) The dynamical state remains unknown. The \citet{McArthur:2014} solution does not support MMR, whereas later analysis by \citet{Rein:2015b} found MMR solutions.
(13) Both resonant and non-resonant configurations are allowed, with the latter slightly preferred \citep{Pepe:2007,Gozdziewski:2007}.
(14) Out of their 850 orbit realizations, 664 turned to be unstable and 28 were in 2:1 MMR. Therefore, the chance to be in MMR is approximately 15\%.
\tablenotetext{a}{We use the parameters from the Keplerian fit (i.e., no planetary interactions) wherever possible. Here the minimum mass $m\sin{i}$ is in unit of $M_{\rm J}$, the orbital period $P$ is in unit of days, and the argument of periapsis $\omega$ in unit of degrees. Under the coplanarity assumption $\varpi=\omega$.}
\tablenotetext{b}{We consider MMR detected if the dynamical analysis (in several cases, strongly) prefers resonant solutions. Otherwise, MMR remains possible although slightly disfavored.}
}
\end{deluxetable}

\end{CJK*}
\end{document}